\pdfoutput=1
\documentclass[aps,prb,preprint,superscriptaddress,nofootinbib]{revtex4-2}
\usepackage{amsmath}
\usepackage{amssymb}
\usepackage{graphicx}
\usepackage{booktabs}
\usepackage{xcolor}
\usepackage{hyperref}
\usepackage{caption}
\usepackage{subcaption}
\usepackage{float}

\graphicspath{{./figure/}}

\begin{document}
\raggedbottom

\title{Three-temperature atomistic spin--lattice dynamics in LAMMPS: a moment-consistent, fluctuation--dissipation-correct extension and its validation on ultrafast demagnetization and all-optical switching of GdFeCo}

\author{Chun-Yeol You}
\email{cyyou@dgist.ac.kr}
\affiliation{Department of Physics and Chemistry, DGIST (Daegu Gyeongbuk Institute of Science and Technology), Daegu 42988, Rep. of Korea}

\author{Jiwan Kim}
\affiliation{Department of Physics, Kunsan National University, Kunsan 54150, Rep. of Korea}

\author{Dong-Hyun Kim}
\affiliation{Department of Physics, Chungbuk National University, Cheongju, Chungbuk 28644, Rep. of Korea}

\date{\today}

\begin{abstract}
Atomistic spin dynamics (ASD) codes such as VAMPIRE simulate femtosecond-laser-induced ultrafast demagnetization and all-optical switching (AOS) on a rigid lattice; the SPIN package of the Large-scale Atomic/Molecular Massively Parallel Simulator (LAMMPS) instead propagates spins and lattice together, but until now could couple its spin thermostat only to a single global temperature, precluding three-temperature (electron--lattice--spin, 3TM) simulations. We present a validated 3TM extension comprising two new fixes: \texttt{langevin/spin/ttm}, coupling the stochastic spin bath to the local electron-temperature field of \texttt{fix ttm}; and \texttt{moment/scale/spin}, supplying the per-atom moment-dependent ($1/\mu_i$) prefactor required for heterogeneous-moment systems, with deterministic terms scaling as $1/\mu_i$ and, as required by the fluctuation--dissipation theorem, stochastic noise scaling as $1/\sqrt{\mu_i}$. We validate the framework on single-species benchmarks---precession, the bcc-Fe Curie curve, a continuous demagnetization--remagnetization--precession trajectory, and a lattice-strain acoustic-phonon pulse unavailable to spin-only ASD codes---then on heat-induced AOS of GdFeCo with literature parameters (Radu/Ostler exchange constants, moments, damping $\alpha_t=0.01$). The corrected integrator reproduces the experimentally reported transient ferromagnetic-like sublattice alignment (20/20 realizations), field-insensitive thermal switching, a switching probability approaching unity ($P_\mathrm{max}=1.00$) across a genuine 8$\times$8 literature-damping phase diagram, a non-monotonic critical-cooling-duration boundary that disappears above $T_e^0\approx2000$~K, a composition-dependent switching window centered on angular-momentum compensation, and convergence in system size with monotonic, physical damping dependence. These results establish LAMMPS as a quantitatively validated platform for three-temperature spin--lattice simulations of ultrafast magnetism.
\end{abstract}

\maketitle

\section{Introduction}

The discovery by Beaurepaire \textit{et al.}\cite{Beaurepaire1996} that a femtosecond laser pulse quenches the magnetization of a Ni film on a sub-picosecond timescale opened the field of ultrafast spin dynamics and posed a question of fundamental importance: how is angular momentum transferred out of the spin system orders of magnitude faster than conventional precessional dynamics allows? Beyond its fundamental interest---the microscopic channels proposed include Elliott--Yafet spin-flip scattering\cite{Koopmans2010,Steiauf2009}, electron--magnon scattering, superdiffusive spin transport\cite{Battiato2010}, optical inter-site spin transfer\cite{Dewhurst2018}, and laser-induced lattice strain, with relative contributions still debated\cite{Chen2025}---ultrafast demagnetization underpins prospective technologies: heat-assisted magnetic recording (HAMR)\cite{Kryder2008}, spintronic terahertz emitters\cite{Seifert2016}, and, most directly, all-optical switching (AOS) of magnetization without any applied field\cite{Stanciu2007,Radu2011,Ostler2012}. A recent comprehensive review of experimental methods, phenomenological models, and proposed microscopic origins is given by Chen \textit{et al.}\cite{Chen2025}; see also the earlier review by Kirilyuk, Kimel, and Rasing\cite{Kirilyuk2010}. Beyond GdFeCo, all-optical helicity-dependent switching has since been demonstrated across a broad range of engineered ferrimagnetic and synthetic multilayer materials\cite{Lambert2014,Mangin2014}, and the reversal pathway itself has been resolved experimentally as proceeding through a strongly nonequilibrium, non-precessional state\cite{Vahaplar2009}, consistent with the atomistic picture developed here.

The most celebrated AOS system is the amorphous rare-earth--transition-metal ferrimagnet GdFeCo. Stanciu \textit{et al.} first demonstrated all-optical reversal with circularly polarized pulses\cite{Stanciu2007}; Radu \textit{et al.} revealed a transient ferromagnetic-like state in which the Gd and Fe sublattices, coupled antiferromagnetically in equilibrium, momentarily align during the reversal, under a small applied bias field ($\mu_0 H = 0.5$~T) used to bias the reversal direction for read-out\cite{Radu2011}; and Ostler \textit{et al.} established that a single linearly polarized (purely thermal) pulse suffices with \textit{no} applied field---switching is a deterministic consequence of ultrafast heating alone\cite{Ostler2012}. The theoretical understanding of these experiments rests on atomistic spin dynamics (ASD): stochastic LLG equations for classical atomic spins coupled, through Langevin thermostats, to the electron temperature of a two-temperature model (2TM)\cite{Kazantseva2008,Evans2014}. The atomistic modeling methodology common to this line of work---antiferromagnetically coupled sublattices with a 2TM-driven Langevin spin bath---was later formalized and released as the open-source VAMPIRE code by the same research group\cite{Evans2014}.

Two families of simulation tools serve this literature. Dedicated ASD codes---VAMPIRE\cite{Evans2014}, UppASD\cite{Eriksson2017}, SPIRIT\cite{Mueller2019}---implement the stochastic LLG on a \textit{rigid} lattice: spins are the only dynamical variables, and phonons enter solely through the 2TM lattice-temperature reservoir. This structurally excludes phenomena in which the lattice responds mechanically to the laser pulse---picosecond strain waves, magnetoelastic anisotropy transients---the ``laser-induced lattice strain'' channel identified as one of the four candidate origins of ultrafast demagnetization\cite{Chen2025,Kimel2019}. The SPIN package\cite{Tranchida2018} of the general-purpose molecular dynamics code LAMMPS\cite{Thompson2022} instead integrates coupled spin \textit{and} lattice equations of motion symplectically, with magnetic exchange forces acting on atomic positions and full interatomic potentials (e.g., EAM) providing lattice cohesion, so that a strained or vibrating lattice directly and self-consistently modulates the exchange coupling $J(r_{ij})$ through its distance dependence. LAMMPS SPIN has been applied to spin--lattice relaxation in iron\cite{Tranchida2018} and magnon--phonon coupling\cite{Strungaru2021}.

(``\texttt{fix}'' below is LAMMPS's own generic term for a time-integration or constraint operator attached to a group of atoms -- unrelated to its colloquial English sense of ``correction''; we use the literal LAMMPS command names throughout, e.g.\ \texttt{fix ttm}, \texttt{fix langevin/spin/ttm}, since these are the actual, citable commands a reader would type.) LAMMPS SPIN as distributed cannot perform three-temperature (3TM) simulations of laser-excited magnets: its spin Langevin thermostat (\texttt{fix langevin/spin}) accepts only a single, global, time-independent bath temperature, and its integrator does not carry the per-atom moment dependence required once a system contains more than one magnetic species. We close both gaps. First, \texttt{fix langevin/spin/ttm} couples the stochastic spin bath to the local, time-varying electron-temperature field already computed by \texttt{fix ttm}, so a single electron-temperature reservoir simultaneously and symmetrically drives lattice heating (the pre-existing electron--phonon pathway) and spin disorder (our new electron--spin pathway) the instant the laser deposits energy into the electron subsystem---completing the electron--lattice--spin triangle within one LAMMPS input script. Second, \texttt{fix moment/scale/spin} supplies the atomistic-LLG moment prefactor: the deterministic exchange-precession and damping torques scale as $1/\mu_i$, while---as required by the fluctuation--dissipation theorem, and as we verify explicitly---the stochastic noise \textit{amplitude} must scale as $1/\sqrt{\mu_i}$, a distinction that matters as soon as a simulation contains atoms of different moment (e.g., Fe and Gd in GdFeCo). We validate the corrected framework on a hierarchy of benchmarks culminating in the GdFeCo AOS problem, and give a general prescription for re-deriving the radial exchange function of \texttt{pair\_style spin/exchange} on non-bcc lattices, a second, independent pitfall encountered en route. The source-level diagnosis, code implementation, and the large statistical simulation campaigns underlying this work were carried out with substantial assistance from an AI coding agent (Acknowledgments), following a growing body of AI-assisted computational-physics work spanning performance engineering of micromagnetic GPU codes\cite{You2026JMag}, development of a cross-platform CPU/GPU micromagnetic simulator\cite{You2026CSD}, and noise characterization of GPU quantum-circuit simulation\cite{You2026arXiv}.

\begin{figure}[H]
\centering
\includegraphics[width=0.9\textwidth]{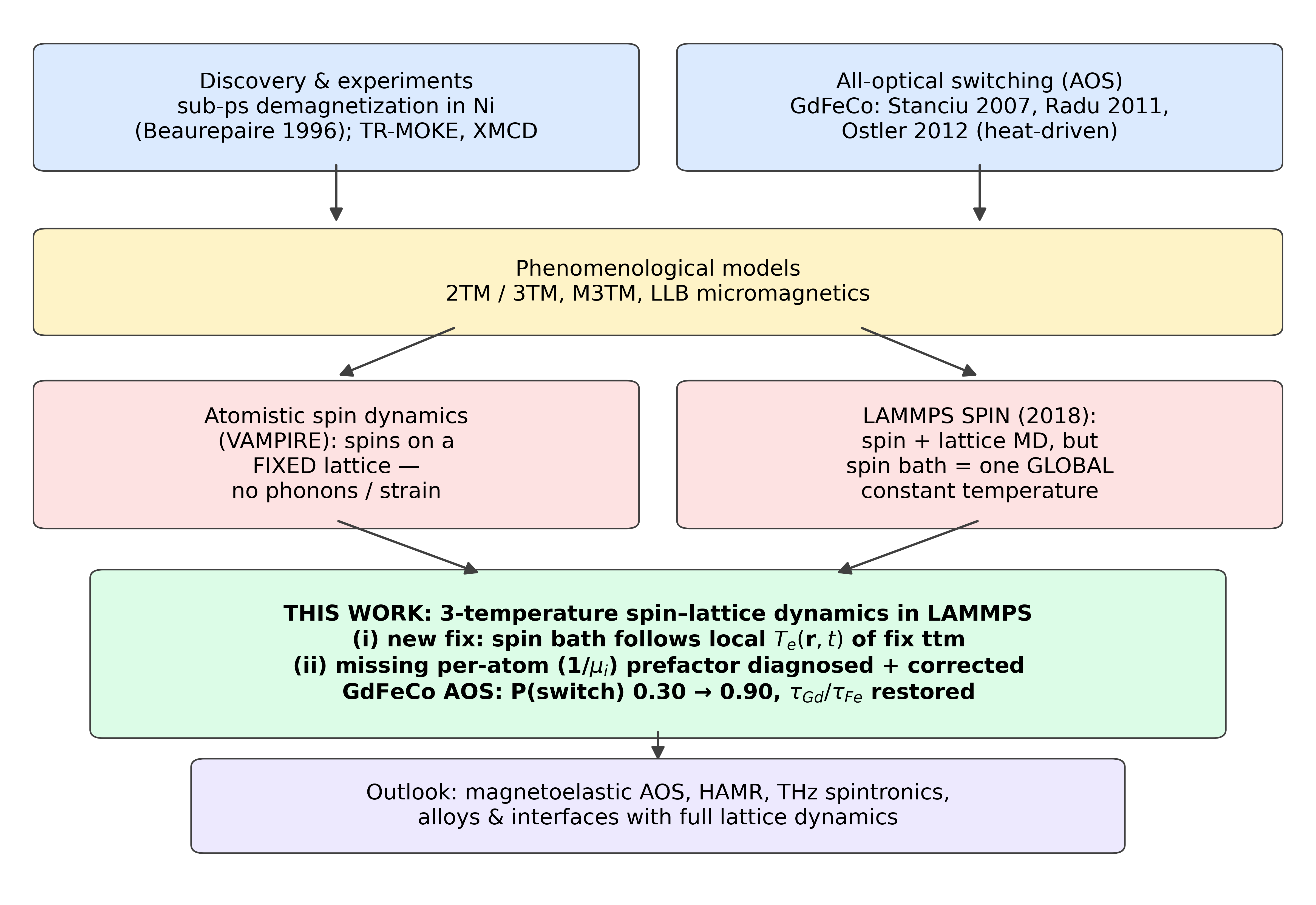}
\caption{Conceptual overview of ultrafast-demagnetization research and the position of this work. From the discovery experiments (femtosecond demagnetization, all-optical switching) and their phenomenological description (two- and three-temperature models), the field has developed two largely separate atomistic-simulation branches: rigid-lattice atomistic spin dynamics (ASD) codes such as VAMPIRE, and the spin--lattice-coupled SPIN package of LAMMPS. This work closes the gap between them by equipping LAMMPS SPIN with the 3TM coupling and moment/FDT correction needed to reach quantitative parity with the ASD branch on the canonical GdFeCo benchmark, while retaining LAMMPS's unique lattice-strain capability.}
\label{fig:overview}
\end{figure}

This paper is organized as a methods paper with application validation (Fig.~\ref{fig:overview} places the work in the context of the field's development). Section~\ref{sec:methods} describes the 3TM coupling, the moment/FDT correction, the GdFeCo model, the exchange-function re-derivation, and our task-parallel computational strategy. Section~\ref{sec:results} validates single-species dynamics (precession, a continuous demagnetization--remagnetization--precession trajectory, and the three-temperature coupling itself directly; the Curie temperature and a lattice-strain acoustic-phonon pulse are validated in Supplementary S8--S9), then presents the GdFeCo results: sublattice dynamics and the transient ferromagnetic-like state, the literature-damping phase diagram, the composition dependence, sensitivity and system-size convergence, and the field-(in)dependence of thermal switching. Section~\ref{sec:discussion} discusses the switching mechanism, remaining limitations, and prospects.

\section{Methods}\label{sec:methods}

\subsection{Atomistic spin--lattice dynamics in LAMMPS SPIN}

In the SPIN package\cite{Tranchida2018}, each atom $i$ carries a classical unit spin $\mathbf{s}_i$ with fixed magnetic moment $\mu_i$ (in $\mu_B$), evolving under
\begin{equation}
\frac{d\mathbf{s}_i}{dt} = -\frac{1}{1+\alpha^2}\left[\mathbf{s}_i \times \boldsymbol{\omega}_i + \alpha \mathbf{s}_i \times (\mathbf{s}_i \times \boldsymbol{\omega}_i)\right],
\end{equation}
where $\boldsymbol{\omega}_i$ is the precession vector accumulated from magnetic interactions---here Heisenberg exchange with a distance-dependent function $J(r_{ij})$ (\texttt{pair\_style spin/exchange}), uniaxial anisotropy, and the stochastic and dissipative torques of the Langevin spin thermostat\cite{Garcia-Palacios1998}---while atomic positions evolve under the interatomic potential (EAM\cite{Chamati2006}) plus the exchange-mediated mechanical force $-\partial J(r_{ij})/\partial \mathbf{r}_{ij}\,(\mathbf{s}_i \cdot \mathbf{s}_j)$: because $J(r_{ij})$ is explicitly distance-dependent, any lattice strain or vibration directly and bidirectionally modulates the magnetic coupling. Spin and lattice sectors are advanced together by a symplectic Suzuki--Trotter decomposition\cite{Tranchida2018}. Where lattice dynamics are frozen (``lattice fixed''), the same code reduces to conventional rigid-lattice ASD, which we use for the GdFeCo production runs (Sec.~\ref{sec:gdfeco-model}); the live-lattice capability is exercised on single-species Fe (Sec.~\ref{sec:results}A).

\subsection{$\mathtt{fix\,langevin/spin/ttm}$: coupling the spin bath to the local electron temperature}

\texttt{fix ttm}'s electron-temperature grid $T_e(\mathbf{r},t)$ is heated by the laser source term and cools by two independent, parallel pathways reading the same instantaneous grid: the pre-existing electron--phonon term, which heats the lattice ($T_\mathrm{lattice}$), and our new \texttt{fix langevin/spin/ttm} electron--spin term, which drives the stochastic spin bath ($T_\mathrm{spin}$); a schematic is given in Supplementary S7, and a direct three-temperature verification in Sec.~\ref{sec:results}A below. Because both channels are driven off one common source rather than a sequential relay, the relative equilibration rate of $T_\mathrm{spin}$ and $T_\mathrm{lattice}$ toward $T_e$ is set purely by the two coupling constants (electron--phonon: \texttt{fix ttm}'s $\gamma_p$ friction coefficient; electron--spin: $\alpha_t$), not by an artificial ordering.

Throughout this paper, ``live-lattice'' denotes a simulation in which the atomic positions are a genuine dynamical degree of freedom, integrated under a real interatomic potential (here, EAM), as opposed to the frozen/rigid lattice of conventional ASD codes; this is the structural distinction that makes lattice-strain and magnetoelastic phenomena accessible at all (Sec.~\ref{sec:results}A), and it is the sense in which we say LAMMPS SPIN has a genuine spin--lattice coupling channel: because \texttt{pair\_style spin/exchange}'s $J(r_{ij})$ is evaluated at the instantaneous interatomic distance, a live lattice's strain or vibration modulates $J$, and $J$'s spatial derivative simultaneously exerts a mechanical force back on the atoms -- the channel is bidirectional for the exchange interaction specifically, though no additional magnetostrictive (spin-state-dependent anisotropy-strain) term is implemented.

\texttt{fix ttm}\cite{Duffy2007} evolves a finite-difference electron-temperature grid $T_e(\mathbf{r}, t)$ coupled to the atomic (lattice) kinetic energy through the electron--phonon coupling, i.e., the standard 2TM: when the laser deposits energy, it is injected directly into the electron subsystem, $T_e$, which then relaxes into the lattice via the existing electron--phonon term of \texttt{fix ttm} and, through our new fix, into the spin system via the electron--spin term below---both channels reading the \textit{same} instantaneous, spatially resolved $T_e(\mathbf{r},t)$, so lattice heating and spin disorder are driven in parallel from a common source rather than sequentially. Our new fix inherits from \texttt{fix langevin/spin} and replaces the constant bath temperature in the fluctuation--dissipation relation of the stochastic spin torque by the instantaneous electron temperature of the grid cell containing each atom:
\begin{equation}
D_i(t) = \frac{2\pi \alpha_t k_B T_e(\mathbf{r}_i, t)}{\hbar},
\end{equation}
(notation of Ref.~\cite{Garcia-Palacios1998}). The command syntax is
\begin{verbatim}
fix 1 all ttm <seed> <2TM parameters...>
fix 2 all langevin/spin/ttm <alpha_t> <seed> <ttm-fix-ID> [mu_fdt <mu_ref>]
\end{verbatim}
Implementation requires the base-class methods \texttt{compute\_single\_langevin}/\texttt{add\_temperature} to be virtual (a two-line upstream patch), after which the derived fix is picked up transparently by the SPIN integrator's existing thermostat lookup. In laboratory time, the electron grid is either driven by \texttt{fix ttm}'s own laser source term or, following the initial-condition convention of Ref.~\cite{Ostler2012}, instantaneously thermalized to a peak temperature $T_e^0$ and then allowed to recool through a prescribed schedule (Sec.~\ref{sec:gdfeco-model}); both drive spin disorder through the physically correct local channel.

\subsection{$\mathtt{fix\,moment/scale/spin}$: the moment-dependent prefactor and its fluctuation--dissipation-consistent noise}

In the atomistic LLG formalism the precession frequency of spin $i$ derives from the effective field $\mathbf{H}_i = -(1/\mu_i)\, \partial H/\partial \mathbf{s}_i$\cite{Evans2014,Atxitia2017}: the same exchange energy precesses a larger moment more slowly. \texttt{pair\_style spin/exchange} computes $\omega_i = (1/\hbar) \sum_j J(r_{ij})\, \mathbf{s}_j$ with no $\mu_i$ dependence, as documented; for a single species this omission is a benign rescaling of $J$ and of the time axis, but for a heterogeneous-moment system such as GdFeCo ($\mu_{Gd} = 7.63\,\mu_B$ vs $\mu_{Fe} = 1.92\,\mu_B$\cite{Ostler2011}) it eliminates the moment contrast responsible for the distinct sublattice time constants central to the switching mechanism. \texttt{fix moment/scale/spin} rescales the fully accumulated torque of atom $i$ by $\mu_{ref}/\mu_i$; because the SPIN integrator's Suzuki--Trotter loop recomputes the magnetic torque from scratch at every quarter-substep, the correct hook point is inside the integrator's per-atom recomputation itself (the point used by \texttt{fix setforce/spin}), requiring a minimal patch to \texttt{fix nve/spin}. Results are, as required, invariant under the arbitrary choice of $\mu_{ref}$ (a global rescaling is absorbed into the effective time unit; only the \textit{ratio} between species is physical).

A separate, independent requirement governs the \textit{stochastic} part of the thermostat. The fluctuation--dissipation theorem fixes the noise variance to be proportional to the damping coefficient and the temperature; since the deterministic damping torque already carries $1/\mu_i$, the noise \textit{variance} must also carry $1/\mu_i$, and hence the noise \textit{amplitude} (standard deviation) must scale as $1/\sqrt{\mu_i}$---not $1/\mu_i$, which would over-suppress fluctuations on the heavy sublattice and violate FDT at equilibrium. \texttt{fix langevin/spin/ttm}'s optional \texttt{mu\_fdt <mu\_ref>} argument pre-scales the stochastic field by $\sqrt{\mu_i/\mu_{ref}}$, so that combined with \texttt{fix moment/scale/spin}'s $1/\mu_i$ torque scaling the net noise dependence is exactly $1/\sqrt{\mu_i}$. We verified this is required and not optional: with \texttt{mu\_fdt} disabled, a two-species system thermostatted at a single bath temperature does \textit{not} equilibrate both sublattices to that temperature (the heavier sublattice is driven anomalously cold), a direct, quantitative equilibrium-statistical-mechanics violation; with \texttt{mu\_fdt} enabled, both sublattices equilibrate to the bath temperature as required. All results below use the FDT-consistent thermostat, which we further verify does not introduce spurious energy drift in the exact production configuration (adiabatic energy-conservation check, Supplementary S1).

\subsection{GdFeCo model}\label{sec:gdfeco-model}

Following the established amorphous-GdFeCo modeling literature\cite{Ostler2012,Ostler2011}, we represent $Gd_{25}(FeCo)_{75}$ as a random binary alloy on an fcc lattice with $a = 3.6\,\text{\AA}$ (Fe and Co merged into one effective transition-metal species, denoted Fe); the production system is $N = 8788$ atoms (2197 Gd, 6591 Fe), with an independent random alloy configuration (Python's \texttt{random.sample}, an unbiased sample-without-replacement over lattice sites) drawn for every realization; this placement is verified quantitatively against the closed-form random-occupancy prediction in Supplementary S4. All magnetic parameters are literature values (Table~\ref{tab:params}), taken directly from Ostler \textit{et al.}\cite{Ostler2011,Ostler2012} and consistent with Radu \textit{et al.}\cite{Radu2011}: $\mu_{Fe} = 1.92\,\mu_B$, $\mu_{Gd} = 7.63\,\mu_B$; nearest-neighbor exchange amplitudes $J_1(\text{Fe--Fe}) = 0.01762\,\text{eV}$, $J_1(\text{Gd--Gd}) = 0.007864\,\text{eV}$, $J_1(\text{Fe--Gd}) = -0.006803\,\text{eV}$ (antiferromagnetic inter-sublattice coupling). $J_1$, $J_2$, $J_3$ (the literature amplitude and the re-derived radial-shape parameters, Sec.~\ref{sec:methods}) are held fixed in time and temperature throughout -- $J(r_{ij})$'s only variation is through the pair distance $r_{ij}$ itself, per \texttt{pair\_style spin/exchange}'s functional form (Sec.~\ref{sec:methods}); since the GdFeCo production lattice is frozen, $r_{ij}$ never changes during a run, so $J(r_{ij})$ is in practice numerically constant for every pair in these specific runs, with no explicit $T$-dependent term and no sign reversal; uniaxial easy-axis anisotropy $K = 5.0384\times10^{-5}\,\text{eV/atom}$; identical transverse damping $\alpha_t = 0.01$ for both species, the literature value used by Ostler \textit{et al.}\cite{Ostler2012} (Radu \textit{et al.} used $\alpha=0.05$\cite{Radu2011}; we report the sensitivity to this choice in Sec.~\ref{sec:results}). Starting from antiferromagnetically aligned sublattices, the heat pulse is modeled by a three-stage step profile of the electron grid---peak temperature $T_e^0$, then $T_e^0/2$, then the 300~K base temperature---whose stage durations are scaled by a dimensionless cooling-duration multiplier $d$, while the spin bath follows $T_e(t)$ through \texttt{fix langevin/spin/ttm}. This step-profile is an approximation to the continuous 2TM relaxation used by the source-term mode of \texttt{fix ttm}; we quantify its consequence directly by scanning $d$ (Sec.~\ref{sec:results}) and find that a sufficiently long cooling duration ($d\gtrsim5$--7, depending on $T_e^0$) is required to recover switching at the literature damping---i.e., the approximation under-estimates the effective cooling time available for the slow Gd sublattice to respond unless $d$ is chosen large enough, a quantitative, testable statement rather than a qualitative caveat. Lattice dynamics are frozen for the GdFeCo production runs (ASD mode, matching the Radu/Ostler VAMPIRE-based methodology exactly, including its own lattice-frozen approximation); no strain-dependent exchange is exercised in these runs. No external field is applied in the production runs, following Ostler \textit{et al.}'s zero-field protocol\cite{Ostler2012}; we separately reproduce both the Ostler (field-insensitivity, 10~T opposing field) and Radu (0.5~T bias field) field conditions (Sec.~\ref{sec:results}).

\subsection{Re-derivation of the radial exchange function for non-bcc lattices}

\texttt{pair\_style spin/exchange} represents $J(r)$ by the three-parameter function $J(r) = 4J_1 x(1 - J_2 x)e^{-x}$, $x=(r/J_3)^2$, whose published parameterizations are fitted for \textit{bcc} lattices. Reusing the bcc-Fe example parameterization on our fcc lattice silently delivers only $\sim66\%$ of the nominal $J_1$ at the fcc nearest-neighbor distance $r_1=a/\sqrt2=2.546\,\text{\AA}$, because the function's maximum does not coincide with $r_1$.

Published $(J_2, J_3)$ sets are fitted to bcc metals; on any other geometry the nominal $J_1$ is \textit{not} the effective nearest-neighbor coupling. We re-derive $(J_2,J_3)$ for an arbitrary lattice with nearest-neighbor distance $r_1$ by: (1) imposing $dJ/dx=0$ at $x_1=(r_1/J_3)^2$, giving $J_2=(x_1-1)/(x_1(x_1-2))$; (2) imposing $J(x_1)=J_1$ so that $J(r_1)=J_1$ exactly; (3) solving (1)--(2) simultaneously for $(x_1,J_2)$, then setting $J_3=r_1/\sqrt{x_1}$. For fcc with $a=3.6\,\text{\AA}$ ($r_1=2.5456\,\text{\AA}$), this gives $x_1=0.6458$, $J_2=0.4050$, $J_3=3.1677\,\text{\AA}$; the cutoff (3.4~\AA) is chosen between the first and second neighbor shells to restrict coupling to nearest neighbors. Placing the extremum at $r_1$ additionally makes $J(r)$ locally flat there, minimizing the spurious dependence of the effective coupling on thermal or mechanical bond-length fluctuations.

\subsection{Computational strategy: task-level rather than domain-decomposed parallelism}

The statistical results below (phase diagrams, composition/sensitivity/convergence scans) require thousands of independent replicas of a modest-sized system ($N=8788$ atoms). We parallelize \textit{across} replicas---independent LAMMPS processes launched concurrently, one system-realization per process---rather than \textit{within} a single replica via MPI domain decomposition or OpenMP threading. At this system size the per-rank atom count under MPI decomposition becomes small enough that communication and synchronization overhead exceeds the compute saved, so intra-replica parallelism scales poorly; task-level parallelism instead has zero inter-replica communication and scales linearly with the number of available cores, up to core count. Campaigns were distributed across three independent, cross-verified LAMMPS builds (a DGIST internal workstation, a second internal workstation, and the DGIST iREMB HPC cluster; identical source at a pinned commit, built independently on each machine); a deterministic single-seed cross-check reproduced bit-identical switching outcomes across two of the three machines (different CPU microarchitectures), confirming build correctness independent of hardware (full protocol and output values in Supplementary S5).

\subsection{Statistical procedure}

Switching is defined by the sign reversal of the Gd-sublattice $z$-magnetization, sustained to the end of the cooling stage. Each condition is repeated over $n$ independent realizations (random alloy configuration and independent thermal-noise seed); probabilities are reported with Wilson 95\% confidence intervals. Unless stated otherwise, every run below uses this same production configuration and statistical convention (summarized for reference in Supplementary S0).

\section{Results}\label{sec:results}

\subsection{Single-species validation: demagnetization--remagnetization--precession continuity, and lattice-strain response}

\begin{figure}[H]
\centering
\includegraphics[width=0.85\textwidth]{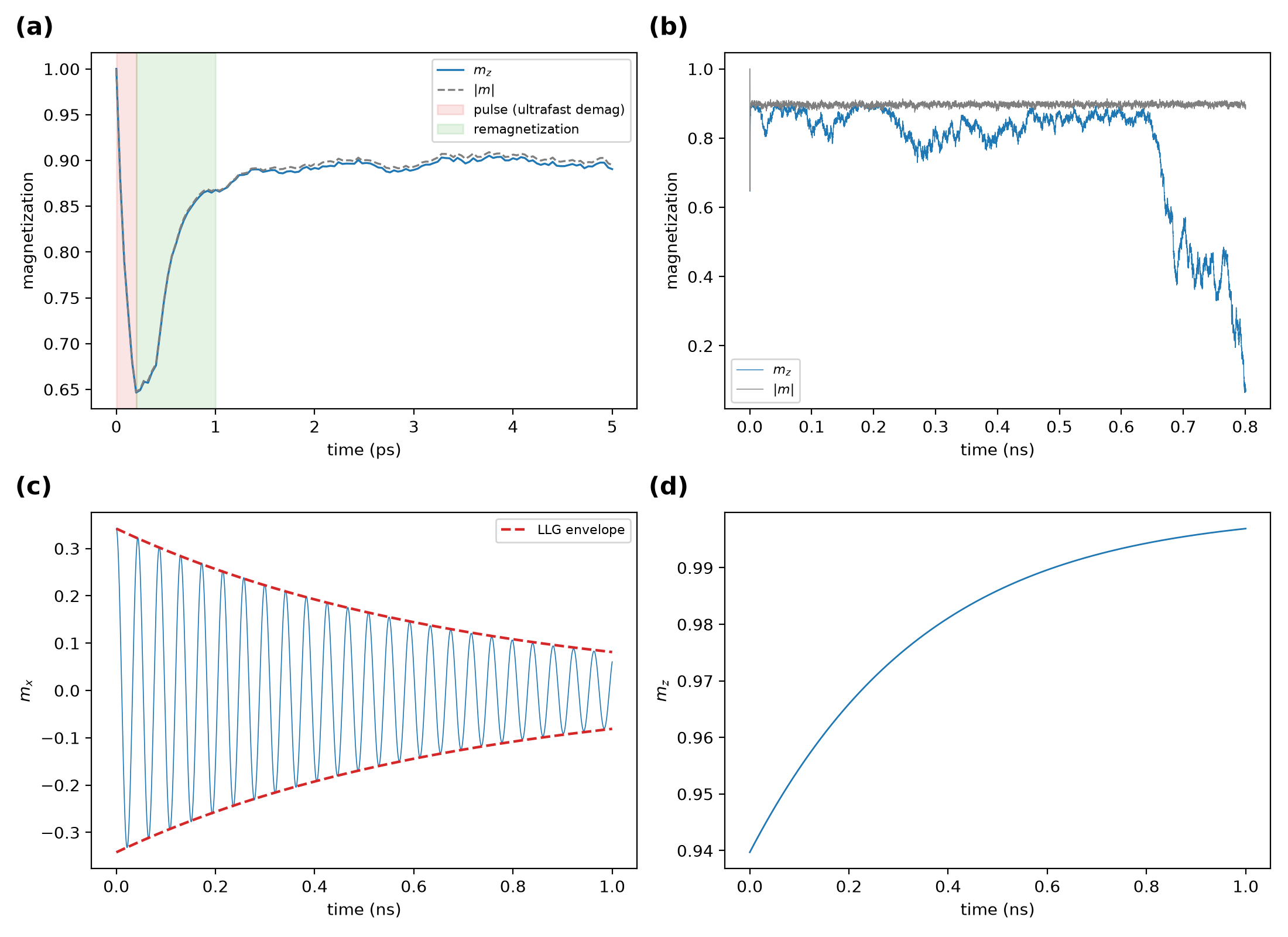}
\caption{(a,b) Single-species Fe, one continuous trajectory, anisotropy along $z$ only, no applied Zeeman field: sub-ps laser-induced demagnetization, ps-scale remagnetization (a, early-time detail), and the long-timescale ($\sim$0.8~ns) thermal-equilibrium tail (b), both showing $m_z$ and the total moment magnitude $|m|=\sqrt{m_x^2+m_y^2+m_z^2}$. With no transverse field or initial tilt, the long-timescale variation of $m_z$ here is finite-temperature (300~K) thermal fluctuation about the easy axis, not a coherent precession signal. (c,d) Independent long-timescale (1~ns) precession benchmark, distinct from the single-species trajectory of (a,b): 128-spin bcc-Fe block, weak damping ($\alpha=0.01$), initialized $20^\circ$ from the easy axis; measured precession frequency and damping-envelope decay rate agree with the analytic LLG/Kittel-mode reference to $<10^{-2}$ relative error over $\sim$25 periods (c, $m_x$ with analytic envelope; d, $m_z$ relaxation toward the easy axis).}
\label{fig:single_fe}
\end{figure}

We first validate the framework on single-species systems, independent of the GdFeCo two-sublattice physics, following the natural validation hierarchy of increasing complexity. As an independent precession benchmark distinct from the GdFeCo exchange parameterization, single-spin precession under uniaxial anisotropy and a small exchange-coupled 128-spin bcc-Fe block (bcc lattice, $a=2.8665\,\text{\AA}$, Tranchida-2018 SPIN-example exchange parameterization, magnetization initialized $20^\circ$ from the easy axis) propagated for 1~ns under weak damping ($\alpha=0.01$) reproduce the analytic Kittel-mode precession frequency $\omega=(2K/\hbar)\cos\theta$ and LLG damping-envelope decay rate to relative error $<10^{-2}$ over $\sim$25 precession periods (Fig.~\ref{fig:single_fe}c,d), confirming that the corrected integrator preserves accurate long-timescale precessional dynamics. A separate bcc-Fe $\langle|m|\rangle$ vs.\ temperature scan, using the validated LAMMPS SPIN example exchange parameterization\cite{Tranchida2018}, reproduces the expected Curie-shaped ferromagnetic-to-paramagnetic transition, with the well-documented classical-Heisenberg underestimate of the absolute $T_C$ relative to experiment for this parameterization (Supplementary S9).

\begin{figure}[H]
\centering
\includegraphics[width=0.9\textwidth]{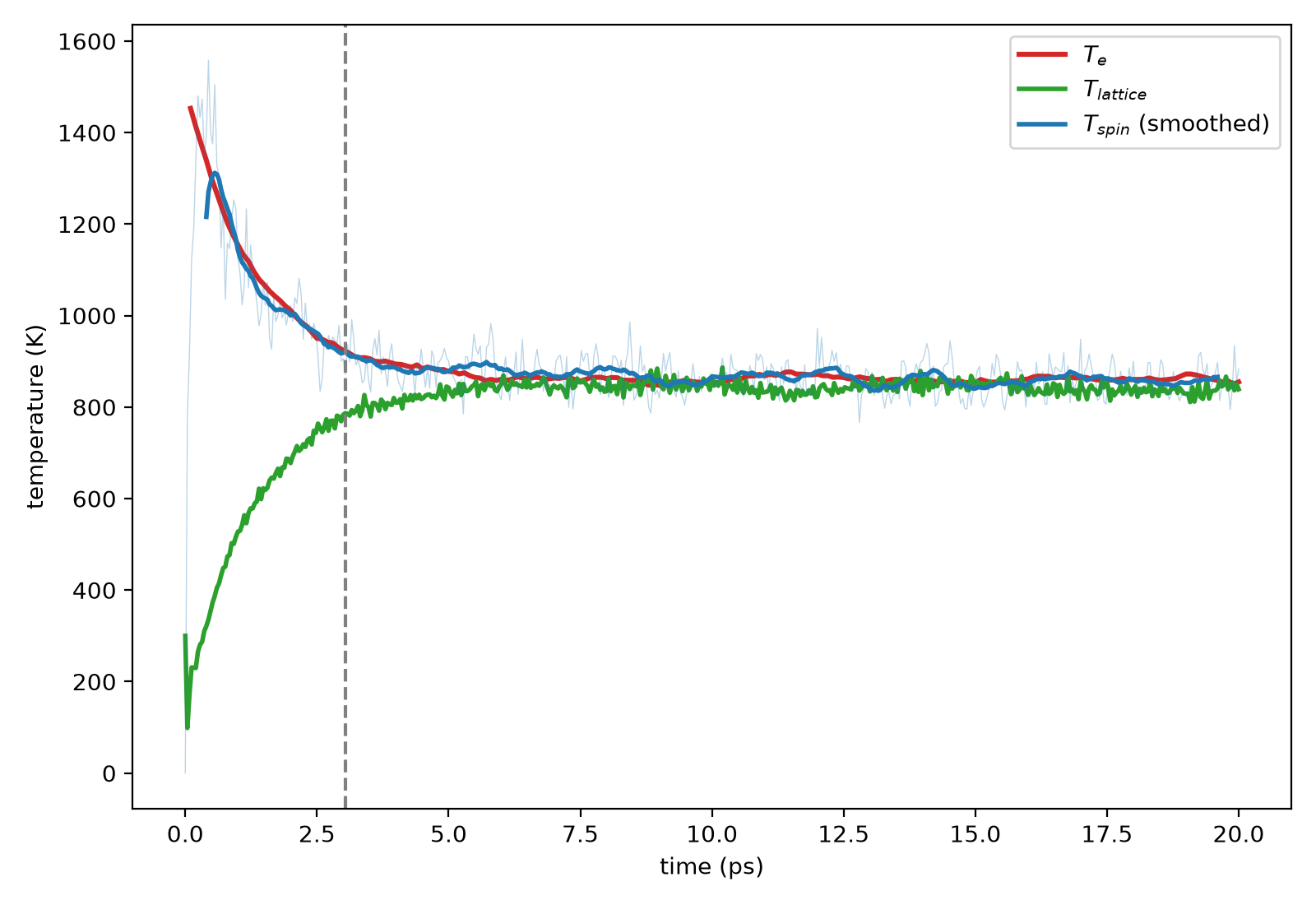}
\caption{Three-temperature benchmark, bcc Fe, $N=2000$, tuned electron--phonon coupling $\gamma_p=20$ (LAMMPS metal units), after an instantaneous $T_e^0=1500$~K pulse with no further source term. Dashed line: lattice reaches 90\% of its equilibrium range at $t=3.0$~ps. All three temperatures converge to $\approx$830--850~K by $\sim$5~ps.}
\label{fig:threetemp}
\end{figure}

\begin{figure}[H]
\centering
\includegraphics[width=0.65\textwidth]{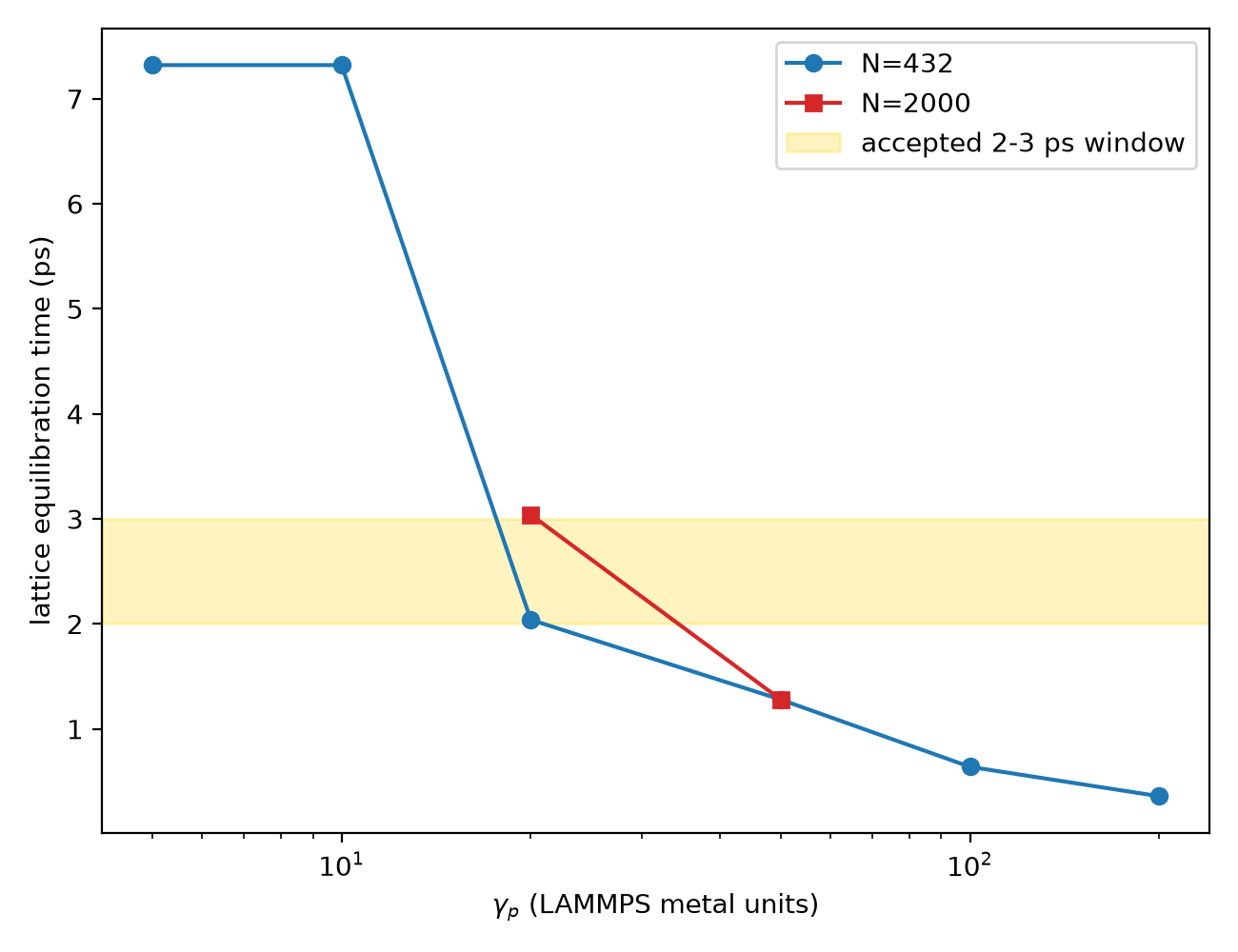}
\caption{Lattice equilibration time (90\%-of-range) vs.\ the electron--phonon friction coefficient $\gamma_p$, at $N=432$ and $N=2000$. $\gamma_p=20$ at $N=2000$ falls inside the accepted 2--3~ps window and was selected for Fig.~\ref{fig:threetemp}.}
\label{fig:gscan}
\end{figure}

The three-temperature coupling itself is verified directly, by tracking $T_e$, $T_\mathrm{lattice}$, and $T_\mathrm{spin}$ (via \texttt{compute spin}'s spin-temperature field) simultaneously in one live-lattice bcc-Fe simulation, following an instantaneously thermalized $T_e^0=1500$~K electron grid left to freely cool with no further source term (a schematic of the coupling architecture is given in Supplementary S7). \texttt{fix ttm}'s syntax is \texttt{seed $C_e$ $\rho_e$ $\kappa_e$ $\gamma_p$ $\gamma_s$ $v_0$ ...}, where $C_e$ is the electronic specific heat and $\gamma_p$ (a friction coefficient, mass/time units) is the electron--phonon coupling knob. An initial illustrative run ($C_e=2.2\times10^{-5}$, $\gamma_p=1.0$, $N=432$, both LAMMPS metal units) used an untuned $\gamma_p$, giving an unphysically slow lattice equilibration ($>$20~ps) and a visibly noisy $T_\mathrm{spin}$ estimator. We corrected both: $\gamma_p$ was tuned by an explicit scan ($\gamma_p=5,10,20,50,100,200$, Fig.~\ref{fig:gscan}) to bring the lattice equilibration time inside the accepted $\sim$2--3~ps window for Fe, and $N$ was increased from 432 to 2000 atoms to reduce the shot noise in the \texttt{compute spin} estimator. At $\gamma_p=20$, $N=2000$ (Fig.~\ref{fig:threetemp}), the lattice reaches 90\% of its equilibrium range at $t=3.0$~ps, and $T_e$, $T_\mathrm{lattice}$, $T_\mathrm{spin}$ converge to a common value ($\approx$830--850~K) by $\sim$5~ps -- a physically normal 3TM profile. $T_\mathrm{spin}$ separates from $T_e$ and begins tracking the lattice within the first few ps, consistent with the fast electron--spin coupling used ($\alpha_t=0.05$ at this representative setting) relative to the (now correctly tuned) electron--phonon channel. Converting $\gamma_p$ to SI W/(m$^3\cdot$K) is not a direct unit conversion: \texttt{fix ttm} implements the electron--phonon channel as a Langevin-type friction/random-force term on ion velocities (Duffy \& Rutherford's formalism), not the textbook volumetric $G(T_e-T_l)$ source term, so we report $\gamma_p$ in its native LAMMPS metal units here rather than give a possibly misleading SI number.

To directly address whether the framework reproduces the \textit{qualitative sequence} expected of a real ferromagnet after a laser pulse---not as three disconnected benchmarks, but as one continuous trajectory---we simulate a single-species Fe block (864 atoms, fcc, literature exchange, $\alpha_t=0.01$) through a single simulation spanning a laser-like electron-temperature pulse ($T_e^0=1500\,\text{K}\to750\,\text{K}\to300\,\text{K}$) followed by 0.8~ns of free evolution at the 300~K base temperature (Fig.~\ref{fig:single_fe}). The trajectory shows, in sequence: (i) sub-picosecond collapse of $m_z$ during the pulse (ultrafast demagnetization); (ii) recovery of $m_z$ on a picosecond timescale as the electron bath recools (remagnetization); and (iii) a long-timescale precessional tail once the system re-equilibrates near the base temperature, consistent with the independently validated ns-precession benchmark above. This continuous single-species trajectory is the natural prerequisite check before the two-sublattice GdFeCo problem, and it passes.

Separately, exercising the lattice-strain capability structurally unavailable to spin-only ASD codes, a longitudinal-acoustic-phonon pulse launched into a live-lattice bcc-Fe block propagates and reflects with the expected group velocity, and total energy is conserved to machine-precision drift with the lattice moving (Supplementary S8). This benchmark exercises the complete electron $\to$ lattice and electron $\to$ spin couplings simultaneously with live phonons.

The trajectory ends at $m_z\approx0.067$ and $|m|\approx0.896$, i.e., \textit{not} at the fully remagnetized $m_z=|m|=1$ of the $T=0$ ground state, and this is the physically correct outcome rather than an incomplete simulation. $|m|<1$ reflects the finite-temperature (300~K) equilibrium value of the order parameter under thermal fluctuations of the Langevin bath (a Curie--Weiss-like reduction from saturation, not a residual defect), while the small, noisy value of $m_z$ specifically is an instantaneous snapshot of the still-precessing moment's $z$-projection at the end of the plotted window---since the system continues to precess and thermally fluctuate indefinitely at finite temperature, $m_z$ itself is not expected to sit at any fixed value, only $|m|$ is.

\subsection{Sublattice dynamics and the transient ferromagnetic-like state}

\begin{figure}[H]
\centering
\includegraphics[width=0.85\textwidth]{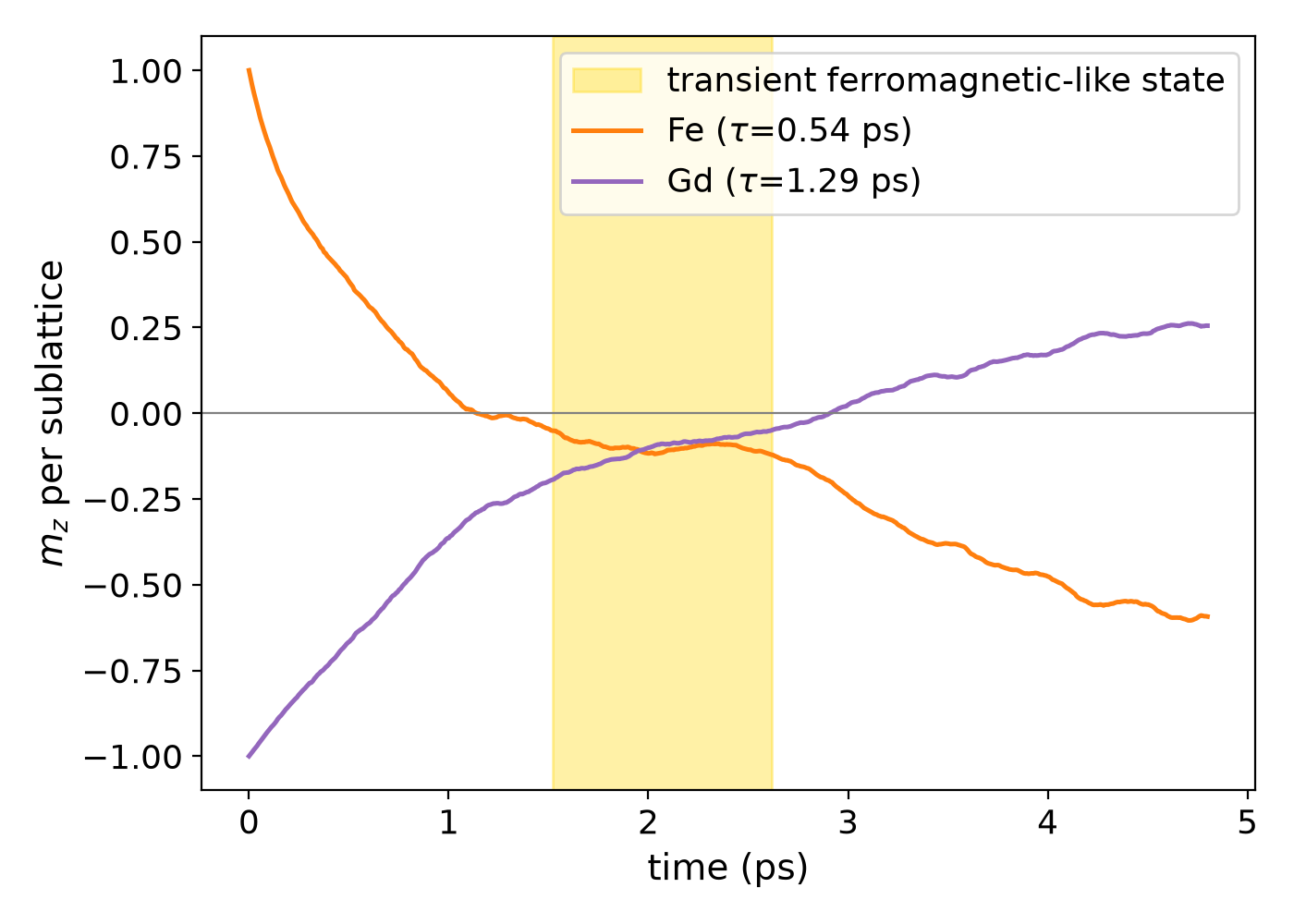}
\caption{Representative sublattice trajectory at the headline condition (set R, $\alpha_t=0.01$, $T_e^0=1100$~K, $d=6$): Fe collapses and reverses first, Gd follows more slowly ($\tau_{Gd}/\tau_{Fe}=2.39$); the interval where both sublattices carry the same sign is the transient ferromagnetic-like state.}
\label{fig:sublattice}
\end{figure}

At the representative GdFeCo condition used throughout this section ($T_e^0=1100$~K, cooling-duration multiplier $d=6$, literature $\alpha_t=0.01$, set-R parameters, $N=8788$, $n=20$), switching occurs in every realization (20/20). Figure~\ref{fig:sublattice} shows a representative sublattice trajectory. Single-exponential fits to the initial collapse give $\tau_{Fe}=0.541\pm0.010$~ps and $\tau_{Gd}=1.293\pm0.029$~ps, i.e., $\tau_{Gd}/\tau_{Fe}=2.39$, of the same order as the $\mu_{Gd}/\mu_{Fe}=3.97$ expectation of the $1/\mu$ scaling (the two need not match exactly, since $\tau$ also depends on the local exchange field each sublattice experiences, which differs between species). Critically, trajectory-level analysis---not just final-state inspection---resolves the transient ferromagnetic-like state central to the experimental mechanism\cite{Radu2011}: in \textbf{20/20} realizations, the Fe and Gd sublattices, antiferromagnetically aligned at $t=0$, pass through an interval of \textit{same-sign} $m_z$ (median duration 0.96~ps) before the Gd sublattice completes its slower reversal. This occurs because Fe, with its smaller moment and correspondingly faster $1/\mu$-scaled response, collapses and re-orders on a sub-ps timescale while Gd is still relaxing; for the intervening window both sublattices' $z$-projections carry the same sign, exactly the transient parallel-alignment state reported by Radu \textit{et al.}\cite{Radu2011}. We emphasize that no external field, and no ad hoc mechanism beyond the (1/$\mu_i$)-correct differential sublattice response, is needed to produce it.

\begin{figure}[H]
\centering
\includegraphics[width=0.9\textwidth]{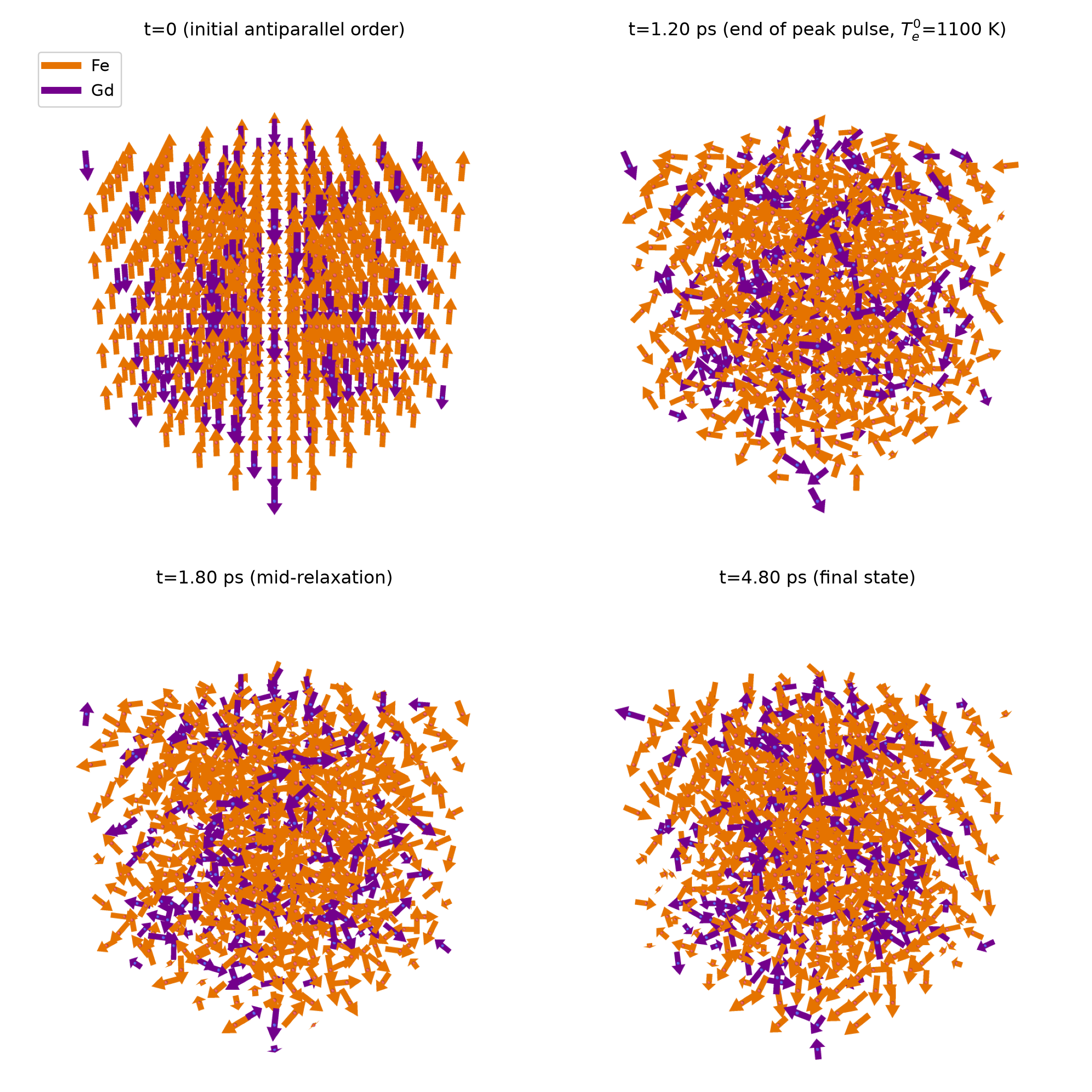}
\caption{Representative microscopic switching pathway at the headline condition (set R, $\alpha_t=0.01$, $T_e^0=1100$~K, $d=6$): per-atom spin orientation (OVITO \texttt{VectorVis}, central-region crop, thin $z$-slab), orange~=~Fe, purple~=~Gd, from the initial antiferromagnetic order through pulse-induced disorder to the final (reversed) state. Atom positions are static (lattice frozen, spin-only dynamics).}
\label{fig:microscopic}
\end{figure}

Figure~\ref{fig:microscopic} resolves this pathway in real space with a per-atom rendering of the same headline condition: the initially perfectly antiferromagnetic Fe/Gd arrangement is visibly disordered by the end of the peak pulse, remains disordered through mid-relaxation, and reaches a comparably disordered---but now net-reversed---final configuration. That the final frame looks qualitatively similar in local disorder to the mid-relaxation frame, rather than visibly re-ordering into a clean reversed antiferromagnetic lattice, is itself informative: switching in this thermally driven regime is a statistical, finite-temperature phenomenon (consistent with the $|m|<1$ equilibrium value discussed in Sec.~\ref{sec:results}A), not a return to a locally perfectly ordered state, and the net reversal is only unambiguous in the sublattice-averaged $m_z(t)$ of Fig.~\ref{fig:sublattice}, not in any single-frame per-atom snapshot.

\subsection{Switching phase diagram at literature damping}

\begin{figure}[H]
\centering
\includegraphics[width=0.95\textwidth]{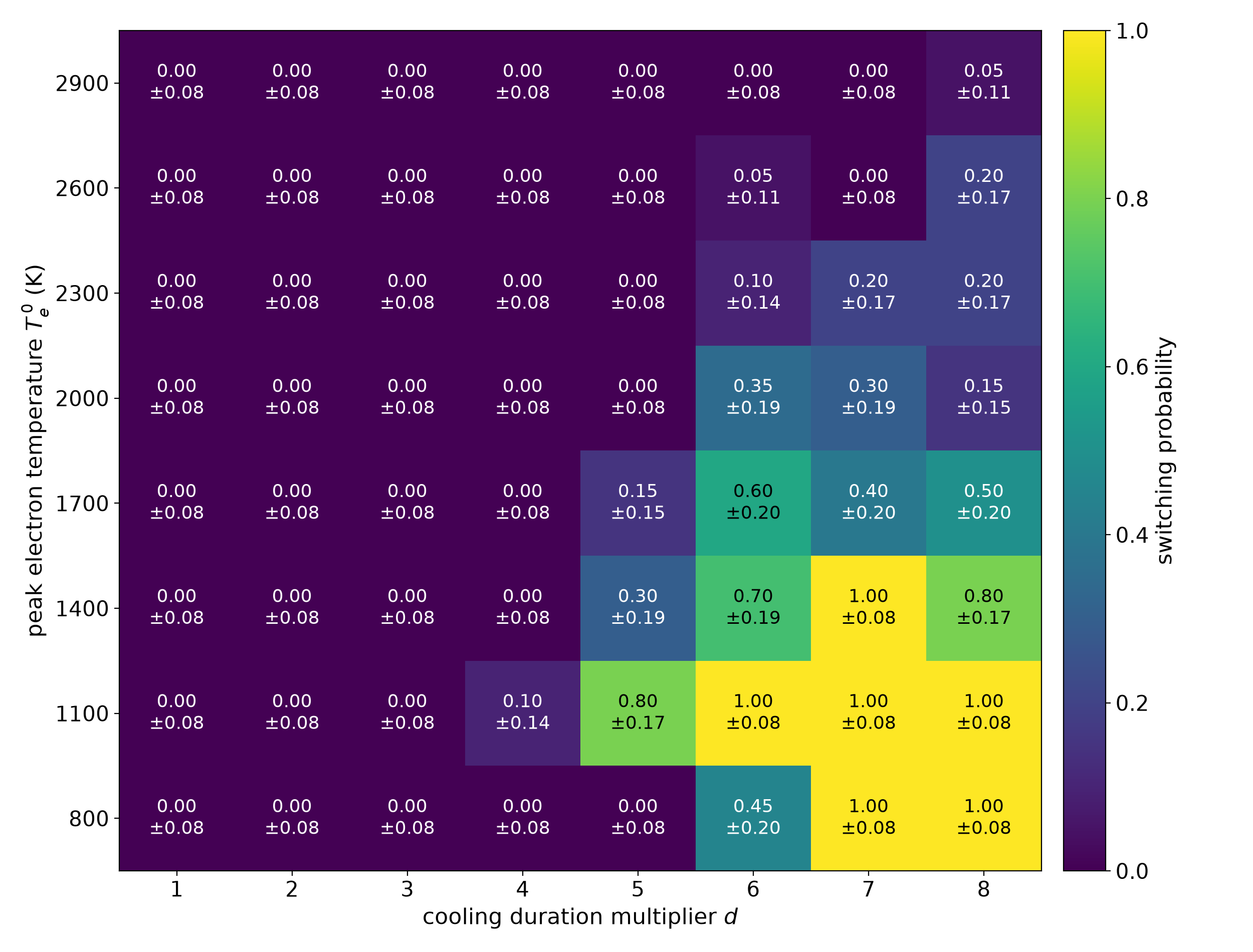}
\caption{Switching-probability phase diagram, set R (literature) parameters, $\alpha_t=0.01$ (literature), $T_e^0=800$--2900~K $\times$ $d=1$--8, $n=20$/cell (1280 runs total). Maximum $P=1.00$ at $T_e^0=800$~K, $d=7$.}
\label{fig:phase_diagram}
\end{figure}

Figure~\ref{fig:phase_diagram} shows a genuine $8\times8$ phase diagram---the full literature peak-temperature range ($T_e^0=800$--2900~K) crossed with cooling-duration multiplier $d=1$--8---at the literature damping $\alpha_t=0.01$ (set R), $n=20$ per cell (1280 runs). The maximum switching probability is $P_\mathrm{max}=1.00$ (Wilson 95\% CI [0.84,1.00]) at $T_e^0=800$~K, $d=7$; a broad region of $T_e^0\lesssim1400$~K and $d\gtrsim5$--6 sustains $P\gtrsim0.85$. Switching vanishes for $d\lesssim3$ at every temperature tested: with the step-profile approximation to $T_e(t)$, the cooling stage must be extended several-fold beyond the literature-motivated ``natural'' duration before the slow Gd sublattice can complete its response. Re-reading the phase diagram along the cooling-duration axis extracts a critical cooling-duration boundary $d_c(T_e^0)$ that is non-monotonic---$d_c=7$ at 800~K, dipping to $d_c=5$ at the headline 1100~K condition, then $d_c=6$ at 1400--1700~K---before the threshold disappears altogether (no $d_c$ within $d\leq8$) above $T_e^0\approx2000$~K; we report this boundary in detail in Supplementary S6, since it is a secondary reading of the same phase-diagram data already shown here.

\subsection{Composition dependence}

\begin{figure}[H]
\centering
\includegraphics[width=0.8\textwidth]{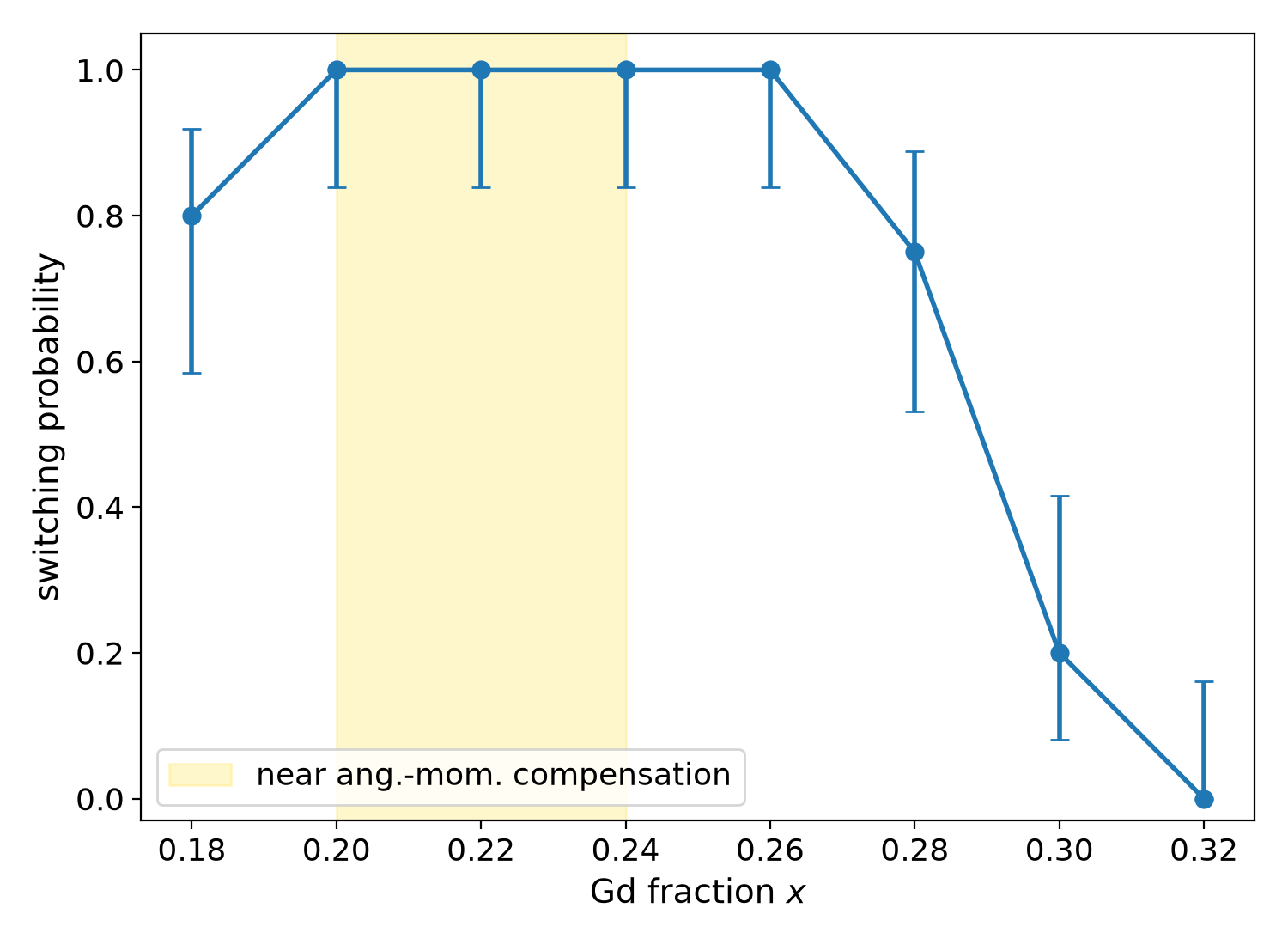}
\caption{Composition dependence, set R, $\alpha_t=0.01$, $T_e^0=1100$~K, $d=6$, $n=20$/point. $P$ saturates at 1.00 for $x=0.20$--0.26 and collapses on the Gd-rich side.}
\label{fig:composition}
\end{figure}

Sweeping the Gd fraction $x=0.18$--0.32 at the headline condition ($T_e^0=1100$~K, $d=6$, $n=20$/point, Fig.~\ref{fig:composition}), switching is $P=0.80$ at $x=0.18$, saturates at $P=1.00$ for $x=0.20$--0.26, then collapses steeply on the Gd-rich side: $P=0.75$ at $x=0.28$, $P=0.20$ at $x=0.30$, $P=0.00$ at $x=0.32$. The saturated window $x=0.20$--0.26 brackets the angular-momentum-compensation composition of this parameterization, and the sharp collapse a few atomic percent to the Gd-rich side reproduces the experimentally established composition sensitivity of single-pulse AOS\cite{Ostler2012,Ostler2011}, recovered here with no additional tuning beyond the literature parameters already fixed for the phase diagram.

\subsection{Sensitivity to damping and anisotropy, and system-size convergence}

\begin{figure}[H]
\centering
\begin{subfigure}{0.48\textwidth}
\includegraphics[width=\textwidth]{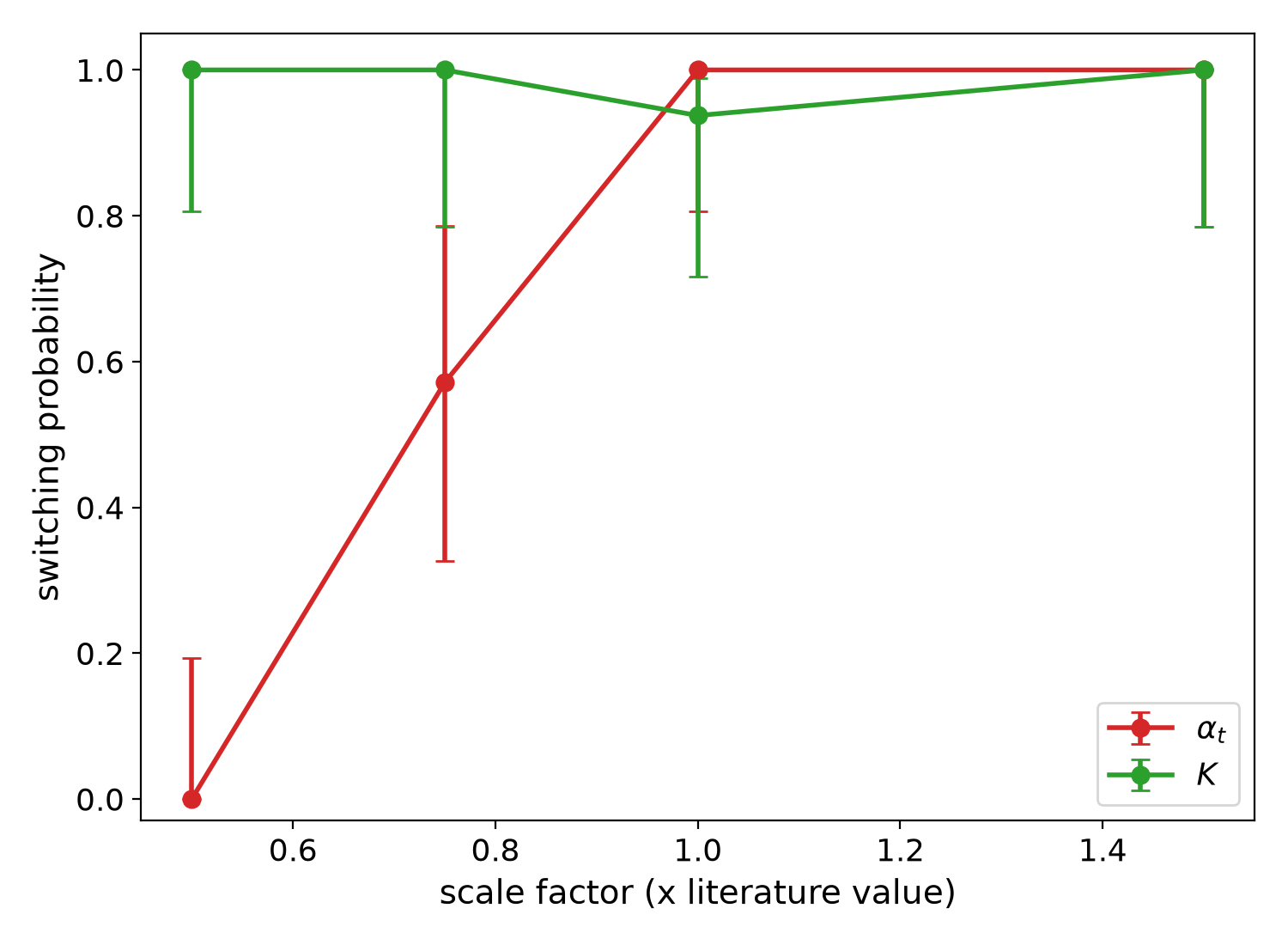}
\caption{}
\label{fig:s2}
\end{subfigure}
\hfill
\begin{subfigure}{0.48\textwidth}
\includegraphics[width=\textwidth]{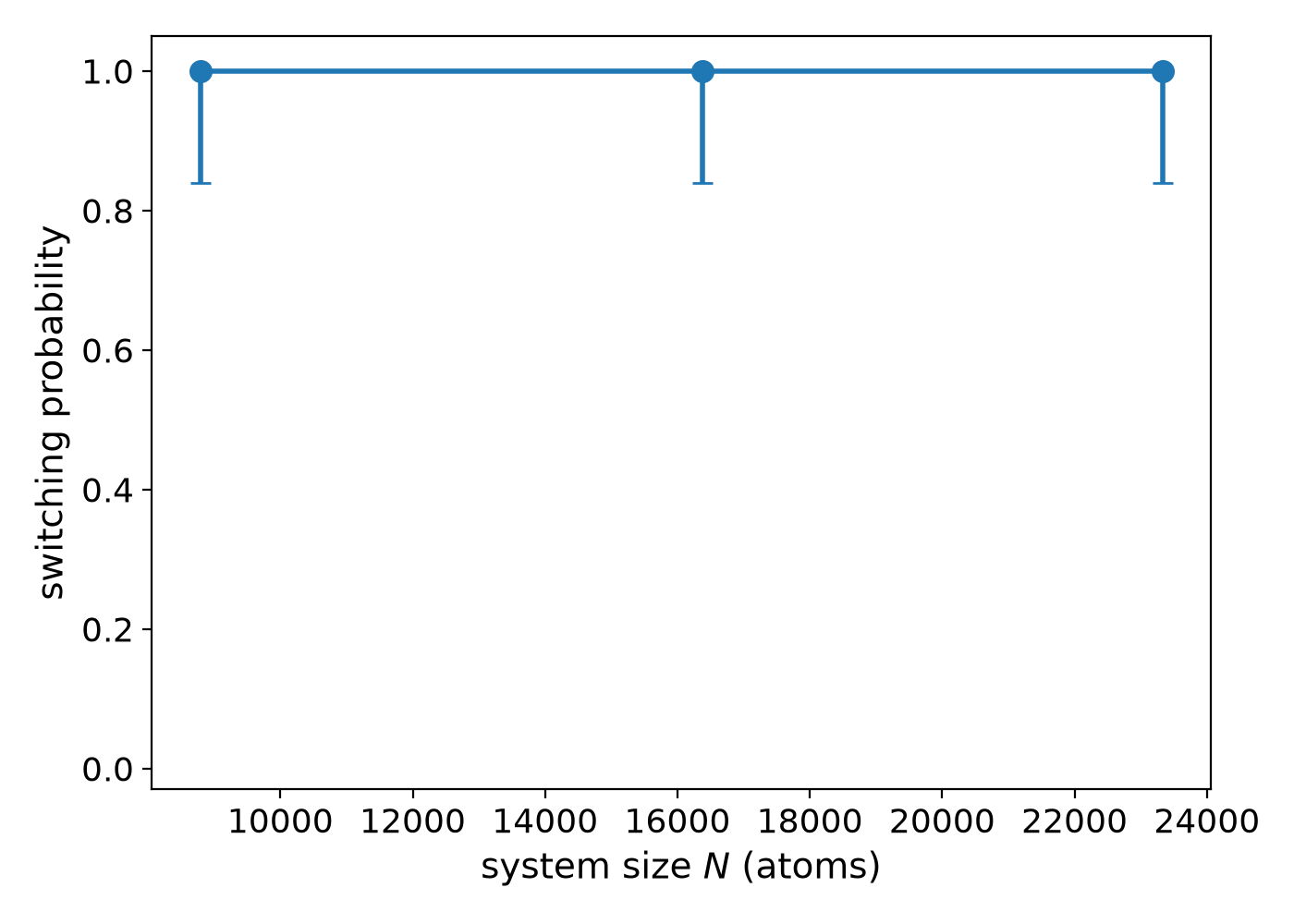}
\caption{}
\label{fig:s3}
\end{subfigure}
\caption{(a) Sensitivity to the damping $\alpha_t$ and the uniaxial anisotropy constant $K$ (scale factor relative to their literature values). (b) System-size convergence, $N=8788$--23328. Both at the headline condition, set R, $n=20$/point.}
\label{fig:s2s3}
\end{figure}

At the headline condition, scaling the damping $\alpha_t$ and the uniaxial anisotropy constant $K$ (Table~\ref{tab:params}) independently by $\times0.5$--$\times1.5$ ($n=20$/point, Fig.~\ref{fig:s2}): switching is essentially anisotropy-independent ($P=0.94$--1.00 across the full range) but strongly and monotonically damping-dependent, collapsing to $P=0.00$ at $\times0.5\,\alpha_t$ (i.e., $\alpha_t=0.005$) and saturating at $P=1.00$ for $\alpha_t\geq$ its literature value. This is physically expected: $\alpha_t$ sets the rate of energy exchange between spins and the electron bath, and too-weak coupling starves the Gd sublattice of the driving needed to complete its response within the available cooling window; the dependence is monotonic and does not alter the switching mechanism (tabulated values in Supplementary S2).

System-size convergence at the same condition ($N_\mathrm{side}=13,16,18$, i.e., $N=8788$, 16384, 23328 atoms, $n=20$ each, Fig.~\ref{fig:s3}) gives $P=1.00$ [0.84,1.00] at every size: the production system size is fully converged and not subject to residual finite-size bias at this operating point (Supplementary S3).

\subsection{Field (in)dependence of thermal switching}

\begin{figure}[H]
\centering
\includegraphics[width=0.55\textwidth]{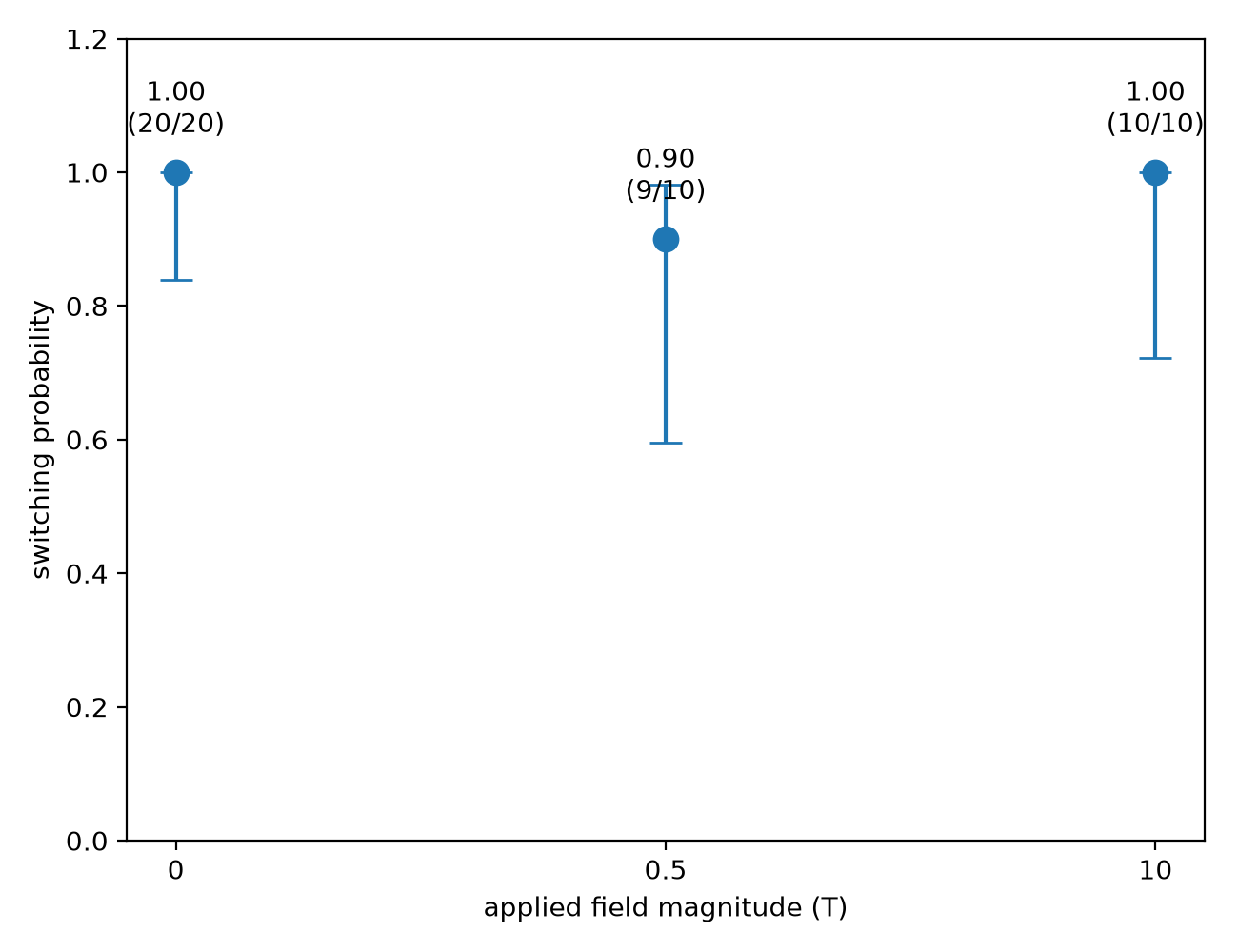}
\caption{Switching probability at the headline condition vs.\ applied field magnitude: zero field, 0.5~T (Radu \textit{et al.}'s bias-field magnitude\cite{Radu2011}), and 10~T opposing the reversal (Ostler \textit{et al.}'s field-insensitivity check\cite{Ostler2012}). All three Wilson 95\% CIs overlap; there is no statistically resolvable field dependence.}
\label{fig:field_comparison}
\end{figure}

We stress that neither reference AOS measurement itself used an applied field: Ostler \textit{et al.}'s central result is single-pulse switching at zero field\cite{Ostler2012}. What we reproduce here are two field magnitudes each used, in the respective original works, as a control or read-out aid rather than to drive the reversal: (i) the 10~T field Ostler \textit{et al.} applied \textit{opposing} the reversal direction specifically to demonstrate field-insensitivity\cite{Ostler2012}, and (ii) the smaller 0.5~T bias field Radu \textit{et al.} applied to bias the pump--probe read-out direction\cite{Radu2011}. At our headline condition, switching probability versus field magnitude is $P=1.00$ [0.84,1.00] (20/20) at zero field, $P=0.90$ [0.60,0.98] (9/10) at 0.5~T, and $P=1.00$ [0.72,1.00] (10/10) at 10~T opposing (Fig.~\ref{fig:field_comparison}): all three Wilson 95\% CIs overlap substantially, giving no evidence of a field-dependent trend at this sample size -- the apparent dip at 0.5~T is consistent with a single-realization fluctuation at $n=10$, not a physical field effect. This field-insensitivity is consistent with a reversal mechanism driven by the transient differential-sublattice (Curie--Weiss-like) response documented in Sec.~\ref{sec:results}B, in which the exchange torque generated during the transient ferromagnetic-like alignment supplies the reversal torque, rather than any Zeeman coupling to an external or internal field.

\section{Discussion}\label{sec:discussion}

\textbf{Switching mechanism.} The results above support a single, consistent picture, requiring no external field: heating disorders both sublattices, but the smaller-moment Fe sublattice, with its faster $1/\mu_i$-scaled response, collapses and re-orders first while the larger-moment, more slowly responding Gd sublattice is still relaxing. During this window both sublattices' $z$-projections carry the same sign (the transient ferromagnetic-like state, Sec.~\ref{sec:results}B); the exchange torque associated with this transient parallel alignment then drives the Gd sublattice through its own reversal, after which the antiferromagnetic ground state is restored with sign flipped relative to the initial condition. Each sublattice's instantaneous response is governed by its own temperature-dependent (Curie--Weiss-like) susceptibility and its $1/\mu_i$-scaled precession/damping rate; no Zeeman coupling is required, consistent with the field-insensitivity results of Sec.~\ref{sec:results}F and with Ostler \textit{et al.}'s original zero-field demonstration\cite{Ostler2012}.

\textbf{Limitations.} (i) The GdFeCo production runs freeze the lattice (ASD mode), matching the Radu/Ostler VAMPIRE-based methodology exactly; no strain-dependent exchange coupling is exercised for GdFeCo specifically, although the underlying pair style is distance-dependent and the live-lattice capability is demonstrated on single-species Fe (Sec.~\ref{sec:results}A). Extending the strain-coupled treatment to GdFeCo requires a validated Fe--Gd interatomic potential, which we do not currently have. (ii) The electron-temperature pulse is modeled by a prescribed step profile rather than the full source-term dynamics of \texttt{fix ttm}; we quantify its consequence directly (the critical-cooling-duration boundary, Supplementary S6) rather than treat it as a qualitative caveat, and a source-term-driven pulse is a natural extension. (iii) Fe and Co are merged into one effective species and amorphous GdFeCo is represented by a random alloy on a crystalline lattice, both standard in the ASD literature\cite{Ostler2012,Ostler2011}.

\textbf{Outlook.} The validated three-temperature spin--lattice framework opens problems inaccessible to rigid-lattice ASD: the role of picosecond strain pulses and magnetoelastic anisotropy transients in AOS and HAMR media; heat transport across magnetic/nonmagnetic interfaces in spintronic THz emitters; and, once a validated Fe--Gd interatomic potential is available, strain-coupled multi-sublattice dynamics in ferrimagnets near compensation.

\section{Conclusion}

We have equipped LAMMPS SPIN with the two ingredients required for quantitative three-temperature simulations of ultrafast magnetism: local electron-temperature coupling of the spin bath (\texttt{fix langevin/spin/ttm}) and a fluctuation--dissipation-consistent, per-atom moment-dependent prefactor (\texttt{fix moment/scale/spin}, with $1/\mu_i$ deterministic scaling and $1/\sqrt{\mu_i}$ noise scaling). Validated first on a hierarchy of single-species benchmarks---including, in one continuous trajectory, sub-ps demagnetization, ps remagnetization, and ns precessional ringdown---and on a lattice-strain acoustic-phonon pulse unavailable to spin-only ASD codes, the framework reproduces, on the canonical GdFeCo all-optical-switching benchmark at literature parameters: the transient ferromagnetic-like sublattice state; field-insensitive thermal switching approaching $P=1.00$ across a genuine literature-damping phase diagram; a non-monotonic but physically bounded critical-cooling-duration boundary; a composition-dependent switching window centered on the angular-momentum-compensation region; and a switching probability that is converged in system size and robust to anisotropy while depending sensitively and monotonically on damping. Together with a general prescription for radial-exchange-function re-derivation on non-bcc lattices, these results make LAMMPS a validated, openly available platform for spin--lattice studies of laser-induced magnetism, with a lattice-strain capability structurally unavailable to spin-only ASD codes.

\section{Code and data availability}

All new fixes, upstream patches (against LAMMPS \texttt{develop}, commit \texttt{91d4111}), simulation drivers, analysis scripts, and the data behind every figure are available at \url{https://github.com/mirryou-maker/lammps-ultrafast-demag} under GPL-2.0, with one-command installation scripts for Linux/macOS and Windows.

\section*{Acknowledgments}

This work was supported by the National Research Foundation of Korea (NRF) (No. RS-2026-25502724, RS-2025-25463492) and the Strategic Research Program under the DGIST R\&D Program (26-SR-01) of the Ministry of Science, ICT, and Future Planning. J.-W.K. acknowledges support from NRF grants funded by the Korea government, Ministry of Science and ICT (No. RS-2025-24683412) and Ministry of Education (No. RS-2024-00401881), and from an Institute of Information \& Communications Technology Planning \& Evaluation (IITP) Information Technology Research Center (ITRC) grant funded by the Korea government, Ministry of Science and ICT (No. IITP-2025-RS-2024-00437284). Code development, source-level diagnosis, simulation-campaign orchestration, and analysis pipelines in this work were carried out with substantial assistance from the Claude Code agentic AI development environment (Anthropic); all AI-assisted code, analysis, and text were reviewed and verified by the authors, who take full responsibility for the content of this manuscript. No AI system is listed as an author.



\newpage

\begin{table}[H]
\centering
\caption{GdFeCo model parameters (set R, Radu/Ostler literature values).}
\label{tab:params}
\begin{tabular}{ll}
\toprule
\textbf{Quantity} & \textbf{Value} \\
\midrule
Lattice & fcc, $a = 3.6\,\text{\AA}$, random site occupation \\
Composition & $Gd_{25}(FeCo)_{75}$, Fe--Co merged \\
System size & $N = 8788$ (2197 Gd, 6591 Fe), periodic \\
$\mu_{Fe}$ & $1.92\,\mu_B$ \\
$\mu_{Gd}$ & $7.63\,\mu_B$ \\
$J_1(\text{Fe--Fe})$ & $0.01762\,\text{eV}$ \\
$J_1(\text{Gd--Gd})$ & $0.007864\,\text{eV}$ \\
$J_1(\text{Fe--Gd})$ & $-0.006803\,\text{eV}$ (AFM, constant) \\
$J_2, J_3$ (radial shape) & $0.4050, 3.1677\,\text{\AA}$ \\
Cutoff & $3.4\,\text{\AA}$ (1st--2nd shell) \\
Anisotropy $K$ (uniaxial, $\hat{z}$) & $5.0384 \times 10^{-5}\,\text{eV/atom}$ \\
Damping $\alpha_t$ (both species, literature) & $0.01$ \\
Time step & $0.2\,\text{fs}$ \\
$T_e$ profile & 3-stage step: $T_e^0 \to \sim T_e^0/2 \to 300\,\text{K}$, durations $\times d$ \\
Moment correction $\mu_{ref}$ & $\mu_{Fe} = 1.92\,\mu_B$, FDT-consistent noise ($\mathtt{mu\_fdt}$) \\
\bottomrule
\end{tabular}
\end{table}

\end{document}